\documentclass[11pt]{article}
\usepackage{amssymb}
\usepackage{enumerate}
\usepackage{MnSymbol}
\usepackage{bbm}
\usepackage{url}
\usepackage{natbib}
\usepackage{graphicx,subcaption}
\usepackage{caption}
\usepackage{float}
\usepackage{morefloats}
\usepackage{tabularx}
\usepackage{setspace,graphicx,amsmath,amsthm,amsfonts,amssymb,mathrsfs}
\usepackage{mathtools}
\usepackage[section]{placeins}
\usepackage{rotating}
\usepackage{epsfig}
\usepackage{times}
\usepackage{enumerate}
\usepackage{array}
\usepackage{siunitx}
\usepackage{booktabs}
\newcolumntype{L}[1]{>{\raggedright\let\newline\\\arraybackslash\hspace{0pt}}m{#1}}
\newcolumntype{C}[1]{>{\centering\let\newline\\\arraybackslash\hspace{0pt}}m{#1}}
\newcolumntype{R}[1]{>{\raggedleft\let\newline\\\arraybackslash\hspace{0pt}}m{#1}}

\usepackage{rotating}
\usepackage{multirow}

\newcommand{\e}{{\rm e}}

\usepackage{indentfirst} 
\usepackage[labelfont=bf,labelsep=period]{caption}   

\usepackage{ulem}
\usepackage{xcolor}

\begin{document}

\author{Sachapon Tungsong, Fabio Caccioli,  Tomaso Aste \\
Department of Computer Science, University College London; \\
Systemic Risk Centre, London School of Economics and Political Sciences}

\title{Relation between regional uncertainty spillovers in the global banking system}

\date{}              

\maketitle
\thispagestyle{empty}

\section*{Abstract}
We report on time-varying network connectedness within three banking systems: North America, the EU, and ASEAN.
The original method by Diebold and Yilmaz is improved by using exponentially weighted daily returns and ridge regularization on vector autoregression (VAR) and forecast error variance decomposition (FEVD).
We compute the total network connectedness for each of the three banking systems, which quantifies regional uncertainty.
Results over rolling windows of 300 days during the period between 2005 and 2015 reveal changing uncertainty patterns which are similar across regions, with common peaks associated with identifiable exogenous events.
Lead-lag relationships among changes of total network connectedness of the three systems, quantified by transfer entropy, reveal that uncertainties in the three regional systems are significantly causally related, with the North American system having the largest influence on EU and ASEAN.

\newpage
\section{Introduction}

Financial markets are increasingly becoming more interconnected \citep{moghadam2010understanding}, and shocks initially affecting one part of the system can quickly propagate to the rest of it. Therefore, understanding the patterns of distress propagation within financial markets is important to characterize systemic risk. 
After the Global Financial Crisis of 2007-2009, significant effort has been devoted into understanding the mechanics of distress propagation within banking systems.
On one hand, a strand of literature focused on modeling the processes through which contagion may occur in interbank networks (see for instance \cite{GlassermanandYoung2016} for a recent review). On the other hand, another strand of literature focused on the quantification of systemic risk from market data (see \cite{AdrianandBrunnerMeier2016, BrownleesandEngle2016}). In particular, \citet{DieboldandYilmaz2009} proposed a method based on Forecast Error Variance Decomposition (FEVD) to estimate from market data networks of interdependencies between firms, and they used the connectedness of the estimated networks to quantify spillovers of uncertainty between variables.

In this paper, we use the methodology of \citet{DieboldandYilmaz2009} -- that we improve through the introduction of an exponential weighting of time series and by coupling it with ridge regression -- to estimate the time evolution of connectedness in three regional banking systems: North America (NA), Southeast Asia (ASEAN), and the European Union (EU). Through VAR and FEVD, we compute the pairwise connectedness between pairs of banks in each region, and we aggregate such pairwise connectedness to compute a measure of total connectedness for the region. 

The time-varying total connectedness computed for each banking system, from a 300 days rolling window during the period 2005-15, indicates temporal changes of systemic risk, with peaks during major crisis events and troughs during normal periods. Analogous results have been observed in other financial systems and different regions \citep{DieboldandYilmaz2009, DieboldandYilmaz2012, DieboldandYilmaz2014, ChauandDeesomsak2014, AlterandBeyer2014, FenglerandGisler2015, Demireretal2015}.
It has to be stressed that, unlike \citet{DieboldandYilmaz2009} who view all financial institutions as belonging to one global system, here we group banks into three regional banking systems. In this way we can perform a comparative analysis between the different regions, which allows us to highlight similarities and differences between them. Furthermore, this allow us to quantify the existence of causal relations between different regions. We must note that combining all the banks together could be somehow misleading because the banks' equities in the three banking systems trade in different stock markets which have significantly different trading hours. 

The main results of our analysis are as follows: First, we notice that the structure of the peaks in the three regional banking systems is very similar with large peaks associated to significant, identifiable major events.
Although the overall patterns are similar, we observe two important differences between the systems. The first is the fact that the overall scale of connectedness is different, with the North-American banking system being more interconnected than the EU, and this being in turn more interconnected than the Southeast Asian system. 
Second, we uncover the existence of lead-lagged relations between the different time series. To quantify this effect, we compute the transfer entropy between the time series associated with changes of connectedness in the different regions, and we uncover the existence of significant net information flows from North America to the EU, from North-America to Southeast Asia, and from the EU to Southeast Asia. 
The robustness of our finding is tested by using different measures for transfer entropy. In particular we find consistent results for the net information flow both with a linear measure of transfer entropy (which corresponds to a Granger causality analysis) as well as with non-linear measures with different parameters. We also retrieve similar causal relation for both one day and five days returns. To the best of our knowledge, this causality study between regional uncertainties is the first of its kind.  

The rest of this paper is organized as follows. 
In Section~\ref{section:literature} we present a literature review and place our paper within the context of previous works.
In Section~\ref{section:data} we describe the used data, while Section~\ref{section:methodology} provides a brief description of our methodology.
Section~\ref{section:results} illustrates and discusses the main results of the paper, and finally we present our conclusions in Section~\ref{section:conclusion}.

\section{Literature review}\label{section:literature}
The literature on spillover effects can be broadly classified into two categories.
The first category comprises network models which aim to describe various causal mechanics of financial contagion, which can be calibrated with balance-sheet data \citep{Furfine2003, DegryseandNguyen2007, UpperandWorms2004, Muller2006, Contetal2010, Upper2011, BirchandAste2014}.
The second category comprises econometric models, which aim at identifying spillover effects exclusively from market data, without making assumptions about the dynamics of distress propagation between banks \citep{AdrianandBrunnerMeier2016, BrownleesandEngle2016}. 
Our paper is close to the second strand of literature, as we try to understand whether market data carry information about the level of interconnectedness between banks, and how exogenous shocks can be amplified by the endogenous dynamics of financial markets.

Network models of contagion go back to the seminal work of \citet{AllenandGale2000}, who showed how the stability of banking system is affected at equilibrium by the pattern of interconnections between banks, and to the work of \citet{EisenbergandNoe2001}, who showed how to consistently compute a clearing vector of payments in a network of interbank claims.
The relation between the structure of an interbank network and its stability has been extensively explored also within the context of non-equilibrium network models (see for instance \citet{Furfine2003}, \citet{Iorietal2006}, \citet{Nieretal2007}, \citet{GaiandKapadia2010}, \citet{Contetal2010}, \citet{Upper2011}, \citet{Battistonetal2012}, \citet{FrickeandLux2015}, \citet{Bardosciaetal2015}), showing in particular the existence of a tension between individual risk and systemic risk -- what makes a bank individually less risky might in fact increase the risk of a systemic failure \citep{Bealeetal2011}.
More recently, these analysis have been extended beyond interbank lending networks to the study of networks of overlapping portfolios \citep{Huangetal2013, Cacciolietal2014, Corsietal2016}. 
Although these models have been insightful to understand the dynamics of financial contagion, and in some cases they have been applied to real data (see \citet{Upper2011} for a review of existing literature), there are clear challenges to their applicability.
First, there is a lack of reliable data on banks' balance sheets, which makes it hard to calibrate models\footnote{\citet{Admatietal2013} report that banks tend to find ways to get around regulations in order to invest in mortgage-backed securities and derivatives via structured-investment vehicles which are off balance sheet items. 
Such leeway allowed by regulations creates regulatory boundaries, making it difficult for outsiders to know what banks actually report.}.
Second, to obtain a reliable assessment of systemic risk one has to capture all relevant types of interconnections between banks as the interaction between different contagion channels can significantly change the stability of the system \citep{Cacciolietal2015}.

Here we take the complementary approach of inferring interdependencies between banks from market data, which belongs to the second strand of literature mentioned above. 
The advantage of the approach with respect to network modeling is that market data are readily available, and that different types of interconnections between banks have already been aggregated by the market. 
The drawback is that this approach does not provide an explanation of how stress propagates between banks, and that it relies on the underlying assumption of market efficiency, which is not realistic \citep{Shiller2003}. 
Nevertheless, one can assume that, although markets are not efficient, prices do reflect to some extent the aggregate information (or expectations) about the underlying assets.
There have been several contributions to this strand of the literature.  
In particular,  \citet{Dungeyetal2005} provide a summary of empirical models of contagion up to 2005. 
 More recent empirical work includes \citet{DieboldandYilmaz2009, DieboldandYilmaz2012, DieboldandYilmaz2014}, \citet{Caceresetal2010}, \citet{Billioetal2012}, \citet{ClaeysandVasicek2014}, \citet{Lucasetal2014}, and \citet{BrownleesandEngle2016}.
Of particular relevance for our paper is the work of \citet{DieboldandYilmaz2009, DieboldandYilmaz2012, DieboldandYilmaz2014} 
which influenced subsequent studies such as \citet{McMillanandSpeight2010}, \citet{Bubaketal2011}, \citet{FujiwaraandTakahashi2012}, \citet{KlobnerandWagner2013}, \citet{AlterandBeyer2014}, \citet{ChauandDeesomsak2014}, \citet{Demireretal2015}, and \citet{FenglerandGisler2015}. 
This strand of contributions uses vector autoregression (VAR) and forecast error variance decomposition (FEVD) to quantify unpredictability of each of the variables in the network. 
By using the VAR and FEVD methods it is possible to disentangle the contribution to unpredictability due to endogenous interdependencies from that due to exogenous shocks. 
Following Diebold and Yilmaz, we will refer to this endogenous component in this paper as total network connectedness, which therefore quantifies the transmission of shocks from banks within the system.

\section{Data}
\label{section:data}
We collect daily stock prices between January 2005 and October 2015 of banks headquartered in North America (US and Canada), the European Union (EU), and Southeast Asia (ASEAN) from Compustat database. 
We select only the financial institutions in the sub-industry ``Banks," i.e., those having GICS code 40101010 and compute log returns from the daily closing prices for each bank.
Our sample includes 10 publicly listed banks in North America, 66 banks in the European Union (EU), and 39 banks in Southeast Asia (ASEAN). 
All banks in the North American banking system have their stocks traded in the New York Stock Exchange (NYSE), while the EU and ASEAN bank stocks mostly trade in their own national stock markets.
Appendix~\ref{appendix:list of banks} provides lists of banks in all three regions as well as their summary statistics.
The data were analyzed over rolling windows of 300 days and over the full period.

\section{Methodology}
\label{section:methodology}

\subsection{Total Connectedness}
Following the approach introduced by \citet{DieboldandYilmaz2009, DieboldandYilmaz2012, DieboldandYilmaz2014} we use a variance decomposition where the forecast error variance of a variable is decomposed into contributions attributed to each variable in the system. 
The approach is based on the vector autoregression (VAR) model, introduced by \citet{Sims1980} (see \citet{StockandWatson2001,Cochrane2005,Lutkepohl2006,Tsay2010} for discussions, reviews and applications). 

VAR estimates the value of a sets of $N$ variables $y_{t,1},...y_{t,N}$ at time $t$ from a linear combination of their values in the past by performing a multi-dimensional regression.
By using the vectorial representation $\mathbf Y_t =  (y_{t,1},...y_{t,N})^T$ and considering the $t-1$ lag only, the regression can be written as: $\mathbf Y_t = \mathbf A \mathbf Y_{t-1}+\mathbf \epsilon_t$ with $ \mathbf A$ an $N\times N$ matrix of coefficients.
By iterating this formula and expressing it in terms of an orthonormal basis of residuals $w_{i,t}$ (with $var(w_{i,t}w_{j,t})=\delta_{i,j}$) \citep{Cochrane2005}, one can write:
\begin{align}
y_{i,t} = \sum_{s=0}^\infty \sum_{j=1}^N \theta_{ij,s} w_{j,t-s} \;\;.
\end{align}
The one-step ahead forecast is $\mathbf{\hat Y}_{t+1} = \mathbf A \mathbf Y_{t}$. 
The forecast error is the difference ${y}_{i,t+1} -{\hat y}_{i,t+1} =  \theta_{ij,0}w_{j,t+1}$ and its variance is therefore:
\begin{align}
var(y_{i,t+1} -  \hat y_{i,t+1}) =  \sum_{j,k=1}^N \theta_{ij,0} \theta_{ik,0} var(w_{j,t+1},w_{k,t+1}) = \sum_{j=1}^N \theta_{ij,0}^2  \;\;.
\end{align}
Each term $\theta_{ij,0}^2$ in the sum is interpreted as the contribution to the one-step forecast error variance of variable $i$ due to shocks in variable $j$.
Its normalized value, $c_{ij}=\theta_{ij,0}^2/\sum_{k=1}^N \theta_{ik,0}^2$, is called connectedness by \citet{DieboldandYilmaz2009, DieboldandYilmaz2012, DieboldandYilmaz2014} and it is associated with the relative uncertainty spillover from variable $j$ to variable $i$.
In this paper we will report about the `total connectedness', which is 
\begin{align}
\mbox{total connectedness} =\frac{1}{N} \sum_{\underset{i\not= j}{i,j=1}}^N c_{ij}
\end{align}
and measures the average effect that the variables have on the one-step forecast error variance.  
It is a measure of spillover uncertainty within the entire system. Larger values of total connectedness correspond to unstable periods with strong influences of the variables uncertainties on each other.

We refine the original \citet{DieboldandYilmaz2009, DieboldandYilmaz2012, DieboldandYilmaz2014} methodology by introducing two technical improvements.
The first improvement consists in the use of ridge regularized VAR \citep{tikhonov1963solution, hoerl1970ridge}, which is used to make estimations less sensitive to noise and uncertainty associated with the finite length of time series. Ridge regression introduces a penalty on the square sum of regression coefficients, thus favoring models with smaller coefficients. This improves regression performances especially for systems with a large number of variables where the covariance matrix is nearly singular (see \cite{gruber1998improving} ). In practice, ridge regression consists in adding a diagonal term in the expression for the regression coefficients: $B = (XX' + \lambda I)^{-1} XY'$ with $I$ the identity matrix and $\lambda$ a coefficient that makes the inversion less sensitive to uncertainty over small eigenvalues \citep{tikhonov1963solution}. The parameter $\lambda$ must be chosen with respect to regression performances, it depends on the length of the time series and on their statistical properties. In our case, we used $\lambda = 100$ which we verified being a good compromise value for this dataset and window length $300$ points. 
We verified that results are little sensitive to variations of $\lambda$ in a wide range $[100-1000]$. 
The second technical improvement consists in the use of exponential smoothing to mitigate the effects associated with sensitiveness to large variations in remote observations, \cite{Pozzietal2012}. Exponential smoothing computes weighted averages over the observation window with exponentially decreasing weights, $\exp(-s/\theta)$, assigned to more remote observations (here $s$ counts the number of points from the present). In this paper we use rolling windows of size $300$ days with exponential weights with characteristic length $\theta =100$. The choice of characteristic length equal to a third of windows length was suggested as optimal by \cite{Pozzietal2012}.

\subsection{Transfer entropy and Granger causality}\label{TEsession}

We investigate how uncertainty in one region affects uncertainty in another region by quantifying lead-lag relationships among uncertainty spillovers. To this purpose we compute the transfer entropy associated with the daily and weekly changes in the total connectedness of the three systems.

In this paper we estimate the transfer entropy by using both linear and non-linear approaches.
The transfer entropy $T_{Y \rightarrow X}$ quantifies the reduction of uncertainty on the variable $X$ that is provided by the knowledge of the past of the variable $Y$ taking in consideration the information from the past of $X$.
In terms of conditional entropies it can be written as: 
\begin{align}\label{TE}
T_{Y \rightarrow X}= H(X_t | X_{t-lag}) - H(X_t | X_{t-lag},Y_{t-lag})
\end{align}
 where $X_t $ represents the present of variable $X$ and $X_{t-lag}$ its lagged past. In this paper we report results for one-day lag.
The conditional entropies are defined as $H(A|B)=H(A,B)-H(B)$ with $H(A,B)$ the joint entropy of variables $A$ and $B$ and $H(B)$ the entropy of variable $B$.

For what concerns the computation of these entropies, the linear approach is the standard procedure. It quantifies the additional reduction in the variance of a variable $Y$ provided by the past of variable $X$ and it is directly related with Granger causality (\citet{granger1988some,Barnettetal2009}).
In this linear case, the entropy associated with a set of variables $Z$ is proportional to the log determinant of the covariance: $H(Z) = \frac{1}{2}\log  \det(2 \e \pi \Sigma(Z))$, where $\Sigma(Z)$ is the covariance matrix of the variables in $Z$. 
By using Eq.\ref{TE} it results that $T_{Y \rightarrow X}$ is simply given by half the logarithm of the ratio between the regression error of variable $X$ regressed with respect to $X_{t-lag}$ and the regression error of variable $X$ regressed with respect to both $X_{t-lag}$ and $Y_{t-lag}$.
The non-linear approach estimates instead entropies by first discretizing the signal into three states, associated with a central band of values within $\delta$ standard deviations from the mean and two external bands respectively with values smaller or larger than the central band. 
By calling $p_A^0$, $p_A^-$ and $p_A^+$ the relative frequencies of the observations in the three bands, entropy is estimated as 
$H(A) = - p_A^-\log p_A^- - p_A^0\log p_A^0 - p_A^+\log p_A^+$. 
The joint entropies are equivalently defined by the joint combination of values of the variables in the 3 bands and the transfer entropy is retrieved by applying Eq.\ref{TE}.

The information flow can be measured by comparing transfer entropies in the two directions. 
If $T_{Y \rightarrow X} > T_{X \rightarrow Y}$, then one can say that the direction of the information goes prevalently from $Y$ to $X$; conversely, if $T_{X \rightarrow Y} > T_{Y \rightarrow X}$, then the direction of the information goes prevalently from $X$ to $Y$.
The net information flow between $X$ and $Y$ can be quantified as $T_{X \rightarrow Y} - T_{Y \rightarrow X} $.

We validated the statistical significance of transfer entropy by comparing our results with the null hypothesis generated by computing 10,000 values of the transfer entropy obtained by randomizing the order of the lagged variables. 
This provides a non-parametric null hypothesis from which p-values can be computed. 
We also compared this non-parametric p-value estimates with the one from F-statistics in the linear case and found comparable results.

\section{Results}
\label{section:results}
\subsection{Total connectedness}
\begin{figure}[H]
\centering
\includegraphics[trim=1cm 0.5cm  1cm 0cm, clip=true, scale=0.5, angle=0]{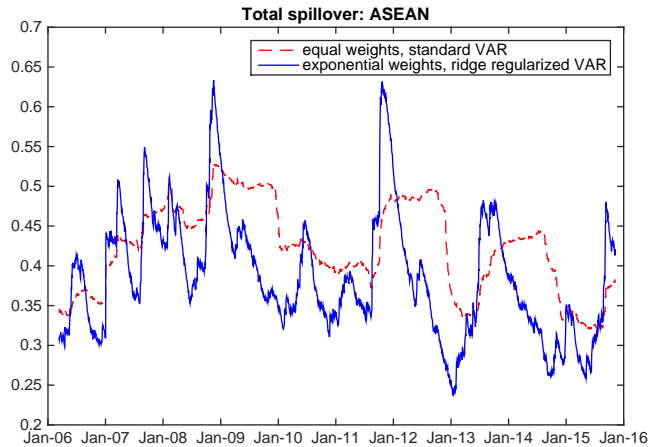}
\caption{ASEAN banking system: Comparison between total connectedness computed with classical VAR approach (dashed red line) and the proposed approach (solid blue line) with ridge penalization and exponential smoothing. 
Computations are over 300-day rolling window.}
\label{fig:ASEAN total connectedness}
\end{figure}

\begin{figure}[H]
\centering
\includegraphics[trim=1cm 0.5cm  1cm 0cm, clip=true, scale=0.5, angle=0]{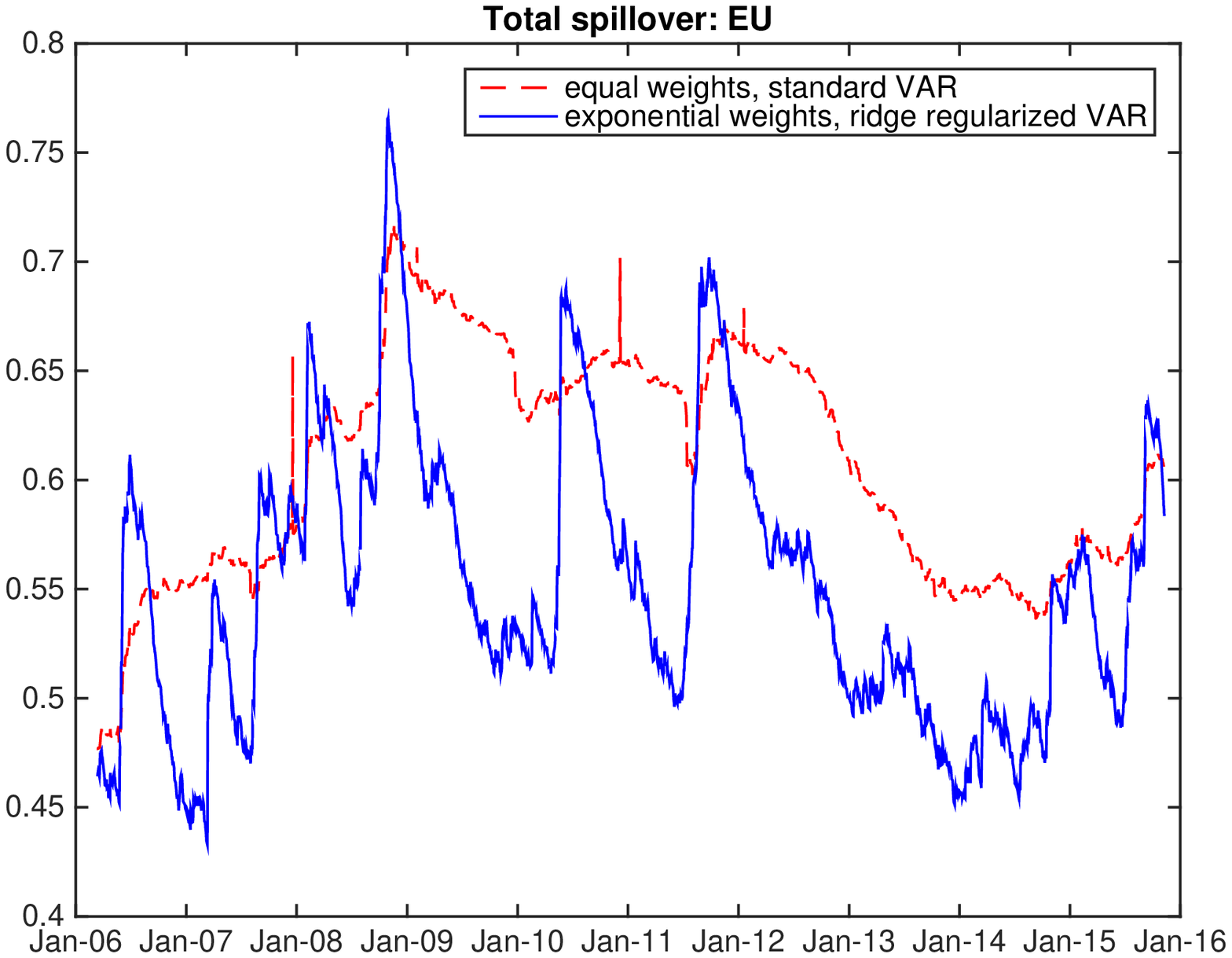}
\caption{EU banking system: Comparison between total connectedness computed with classical VAR approach (dashed red line) and the proposed approach (solid blue line) with ridge penalization and exponential smoothing. 
Computations are over 300-day rolling window.}
\label{fig:EU total connectedness}
\end{figure}

\begin{figure}[H]
\centering
\includegraphics[trim=1cm 0.5cm 1cm 0cm, clip=true, scale=0.5, angle=0]{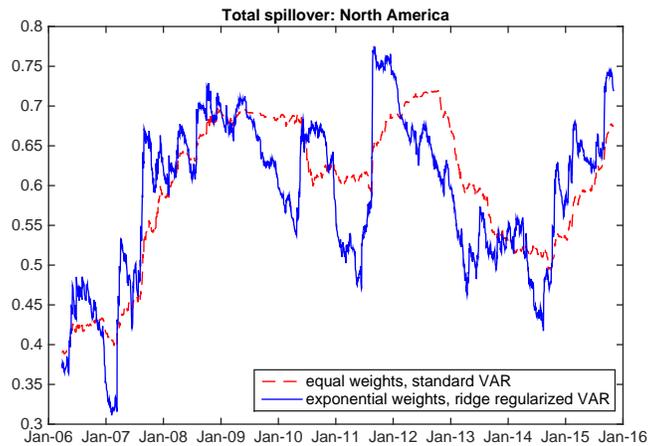}
\caption{North American banking system: Comparison between total connectedness computed with classical VAR approach (dashed red line) and the proposed approach (solid blue line) with ridge penalization and exponential smoothing. 
Computations are over 300-day rolling window.}
\label{fig:NA total connectedness}
\end{figure}

We compute the total connectedness of the three banking systems -- North America, EU and ASEAN -- over a rolling window of 300 days for the ten years period between January 2006 and January 2016.
Figures~\ref{fig:ASEAN total connectedness}, ~\ref{fig:EU total connectedness}, and ~\ref{fig:NA total connectedness} report the results for each of the three systems comparing the original approach of \citet{DieboldandYilmaz2009} (in dashed red line) with the improved approach proposed in this paper (solid blue line).
Let us first observe that the total connectedness from the two approaches have similar values and comparable behavior. 
We can observe that the use of ridge regularized VAR eliminates some outlying spurious peaks observed with the original method. For instance, this is particularly evident in Fig.\ref{fig:EU total connectedness} for the peak after January 2011, but the effect is present in all samples across the three regions and periods.
More evident is the effect of exponential smoothing, which makes peaks sharper and eliminates the plateau effect due to the persistence of the influence of a peak during the whole length of the rolling window. For instance, this is especially evident in Fig.\ref{fig:ASEAN total connectedness} where in the standard VAR method the peak in total connectedness observed just after January 2009 persists creating a plateau that drops abruptly after 300 days in January 2010. 
Conversely the exponential weighted ridge regularized method reveal a clear peak reaching maximum around January 2009 followed by a sharp decrease.
We observe that the plateau effects in standard VAR-equal-weights method sometimes hide completely peaks that are instead detected with the exponentially weighted ridge regularized method; this is for instance the case for the late 2010 North-America spillover peak visible in Fig.\ref{fig:NA total connectedness} only in the exponentially weighted ridge regularized method.

Let us note that in \citet{DieboldandYilmaz2009}, where total connectedness in equity index returns and equity index return volatilities were measured, they found that the return spillovers demonstrate ``a gently increasing trend but no bursts, whereas volatility spillovers display no trend but clear bursts." 
Our results in Figures~\ref{fig:ASEAN total connectedness}, ~\ref{fig:EU total connectedness}, and ~\ref{fig:NA total connectedness} indicate that applying exponential weights onto the returns allow us to observe both trends and bursts in the return uncertainty spillovers.


A comparison between ASEAN, EU, and North American total connectedness from the ridge regularized VAR models is presented in Figure~\ref{fig:ASEAN, EU, NA total connectedness}, where major events are labeled on the graph when they occurred.
We observe that the general shapes of the total connectedness of the three banking systems appear to be similar. Over the 10-year period from 2006 to 2015, the values of the North American total connectedness are generally higher than those of the EU and ASEAN banking systems with the exceptions in 2006 to mid 2007, early 2011, early 2013 and mid 2014.
From visual inspection of Figure~\ref{fig:ASEAN, EU, NA total connectedness}, we notice that variations in total connectedness of the North American banking system seems to lead those of the EU and ASEAN systems and total connectedness of the EU system seems to lead that of the ASEAN system.
This prompts us to perform causality tests on the total connectedness time series of the three banking systems in order to investigate how systemic uncertainty in each region influences the others and the lead-lag relationships among them.
\begin{figure}[H]
\centering
\includegraphics[trim=3cm 1cm 3cm 0cm, clip=true, scale=0.50, angle=90]{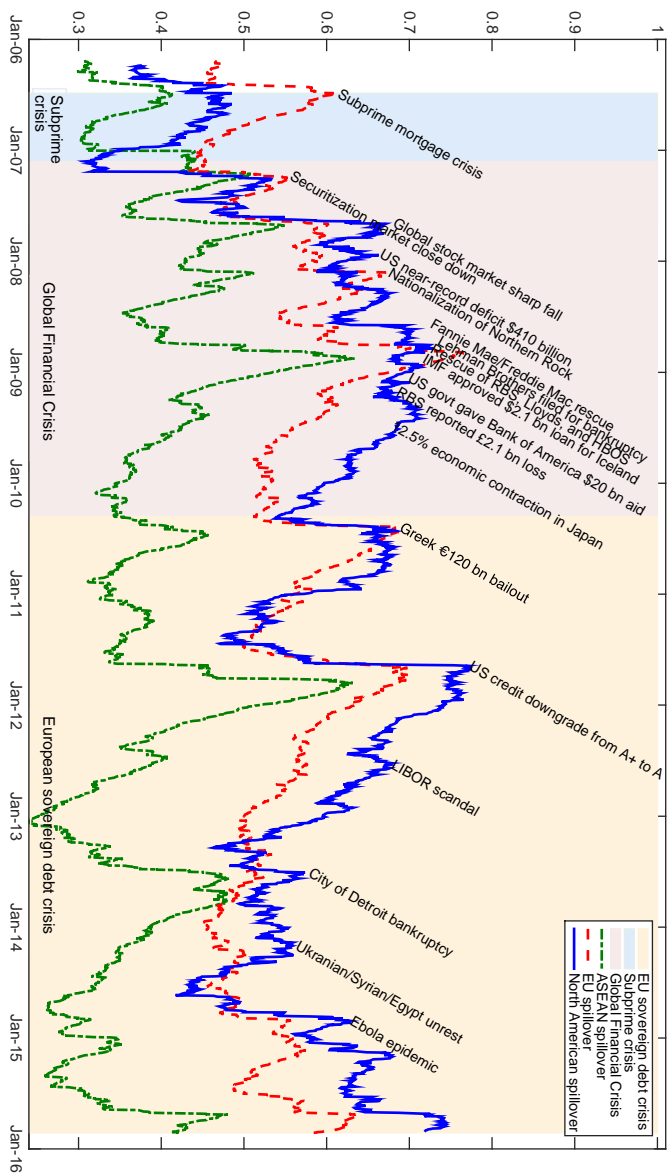}
\caption{Total connectedness in the three banking systems (as in Figs.\ref{fig:ASEAN total connectedness}, \ref{fig:EU total connectedness} and \ref{fig:NA total connectedness}, solid lines). Major events associated with peaks are indicated in the figure.  
Computations are over 300-day rolling window.}
\label{fig:ASEAN, EU, NA total connectedness}
\end{figure}

\subsection{Causality tests on regional total connectedness}
\label{section:Transfer Entropy tests}
In order to quantify the lead-lag relationships among the North American (NA), EU, and ASEAN (AS) total connectedness we compute transfer entropy and information flow between the daily changes of total connectedness in the three regions for one-day lag.
Results are reported in Tab.\ref{table:Transfer Entropy full sample daily returns}. 
Transfer entropies are estimated using both linear and non-linear approaches discussed in section \ref{TEsession}. 
We recall that the linear measure is equivalent to Granger causality, where a significant transfer entropy corresponds to a validated Granger causality relation. 
The non-linear measure are computed for fluctuation bands at $\delta = 1, 2, 3$ standard deviations (see section \ref{TEsession}). 
One can observe that there is a significant information transfer between NA and EU, NA and AS and EU and AS, that for the linear case, implies NA Granger causes EU, NA Granger causes AS and EU Granger cause AS. We observe that the non-linear estimation gives consistent results with the linear estimate for all values of $\delta$, demonstrating robustness of the result. 
We also observe that there are significant causal relations also in the opposite directions.  
Given the extended time-lags between the three regions it is fair to question whether one-day time lag and one-day time horizon will affect asymmetrically markets depending on their relative opening hours. We therefore test the flow of information across regions for time-horizon and lag of 5 days instead of one day. The results for the transfer entropies and information flow, performed for the entire period on non-overlapping 5-day returns, are reported in Tab. \ref{table:Transfer Entropy full sample daily returns 5 days}.    
We observe that results are consistent with the ones for one-day time horizon and lag reported in Tab. \ref{table:Transfer Entropy full sample daily returns} with the main difference being the lower statistical significance. This is expected because the time series for the 5-day changes are five times shorter than the ones for daily changes.

\begin{table}[H]
\centering
\caption{Quantification of transfer entropy between regional total connectedness: March 28, 2006-November 2, 2015 (full sample). From daily changes in the total connectivity using one day lag.}
\label{table:Transfer Entropy full sample daily returns}
\begin{footnotesize}
\begin{tabular}{@{}lllllll@{}}
\toprule
method & $TE_{NA \rightarrow EU}$ & $TE_{EU \rightarrow NA}$ & Net Information Flow\\ \midrule
 linear   				        &      0.004722$^{**}$	   & 	0.001354$^*$ 		& 0.003369	\\
 non-linear treshold $\sigma$   &	0.005251$^{***}$ &	0.006711$^{**}$ 	& -0.001460\\
 non-linear treshold $2\sigma$ &	0.003980$^{***}$ &	0.002012$^*$		& 0.001968\\
 non-linear treshold $3\sigma$ &	0.004939$^{***}$ &	0.000561			& 0.004378\\
\bottomrule
\toprule
method  & $TE_{NA \rightarrow AS}$ &  $TE_{AS \rightarrow NA}$ & Net Information Flow\\ \midrule
 linear   					&      0.017336$^{***}$ 	& 0.008931$^{***}$	& 0.008405	\\
 non-linear treshold $\sigma$ 	&	0.008789$^{***}$	& 0.005837$^{**}$	& 0.002953	\\
 non-linear treshold $2\sigma$ 	&	0.005348$^{***}$	& 0.002305$^{*}$	& 0.003042	\\
 non-linear treshold $3\sigma$ 	&	0.003150$^{**}$	& 0.002803$^{***}$	& 0.000348	\\
 \bottomrule
\toprule
method  & $TE_{EU \rightarrow AS}$ &  $TE_{AS \rightarrow EU}$ & Net Information Flow\\ \midrule
 linear   					&      0.005659$^{**}$		&0.003633$^{**}$	&0.002026	\\
 non-linear treshold $\sigma$ 	&	0.005553$^{**}$		&0.001262		&0.004291	\\
 non-linear treshold $2\sigma$ 	&	00.005960$^{***}$		&0.000228		&0.005732	\\
 non-linear treshold $3\sigma$ 	&	0.004238$^{***}$		&0.002118$^{***}$	&0.002120	\\
 \bottomrule
{\tiny  * p-value $< 0.05$, ** p-value $< 0.01$, *** p-value $< 0.001$.}
\end{tabular}
\end{footnotesize}
\end{table}
\begin{table}[H]
\centering
\caption{Quantification of transfer entropy between regional total connectedness: March 28, 2006-November 2, 2015 (full sample). From weekly changes (5 days) in the total connectivity using five days lag.}
\label{table:Transfer Entropy full sample daily returns 5 days}
\begin{footnotesize}
\begin{tabular}{@{}lllllll@{}}
\toprule
method  & $TE_{NA \rightarrow EU}$ &  $TE_{EU \rightarrow NA}$ & Net Information Flow\\ \midrule
 linear   					&     0.008003$^{*}$	   & 	0.001255 			& 0.006747	\\
 non-linear treshold $\sigma$  &	0.009204		   &	0.009474 			& -0.000271\\
 non-linear treshold $2\sigma$ &	0.017228$^{***}$ &	0.003196			& 0.014032\\
 non-linear treshold $3\sigma$ &	0.024087$^{***}$ &	0.002335$^*$		& 0.021752\\
\bottomrule
\toprule
method  & $TE_{NA \rightarrow AS}$ &  $TE_{AS \rightarrow NA}$ & Net Information Flow\\ \midrule
 linear   					&      0.017200$^{**}$ 	& 0.003703		& 0.013497	\\
 non-linear treshold $\sigma$ 	&	0.010598$^{*}$		& 0.004354		& 0.006244	\\
 non-linear treshold $2\sigma$ 	&	0.006509			& 0.006475		& 0.000034	\\
 non-linear treshold $3\sigma$ 	&	0.002107			& 0.006805$^{***}$	& -0.004698	\\
 \bottomrule
\toprule
method  & $TE_{EU \rightarrow AS}$ &  $TE_{AS \rightarrow EU}$ & Net Information Flow\\ \midrule
 linear   					&      0.022020$^{**}$		&0.000619	&0.021401	\\
 non-linear treshold $\sigma$ 	&	0.021641$^{***}$		&0.002374	&0.019267	\\
 non-linear treshold $2\sigma$ 	&	0.022964$^{***}$		&0.002900	&0.020063	\\
 non-linear treshold $3\sigma$ 	&	0.007488$^{**}$		&0.000405	&0.007083	\\
 \bottomrule
{\tiny  * p-value $< 0.05$, ** p-value $< 0.01$, *** p-value $< 0.001$.}
\end{tabular}
\end{footnotesize}
\end{table}

\section{Conclusion}
\label{section:conclusion}

We investigate regional and inter-regional uncertainty spillovers in the North American, EU, and ASEAN banking systems during a period characterized by great regional and global financial stress (2005-2015).
Uncertainty and financial instability is quantified by means of total network connectedness, that we measure improving the method of Diebold and Yilmaz. 
We demonstrate that exponential smoothing and ridge regression provide better defined peaks in the temporal analysis and avoid the occurrence some spurious peaks. 
We observe that the North-American system appears to be consistently more interconnected than the EU, which in turn is more interconnected than the ASEAN network. Similarly to previous analysis of Diebold an Yilmaz on other systems, our empirical analysis of the North-American, ASEAN and EU banking networks shows that increased connectivity corresponds to periods of higher distress in the system. 
We observe that all large peaks of total network connectedness are associate with identifiable major exogenous events.
Despite some of these events being related to specific regions, the effects are seen similarly across the three banking systems, which reveal similar patterns of peaks and troughs in the variations of their total network connectedness. 
However, such variations are not perfectly synchronous across the regions, and causality patterns are discovered by using transfer entropy.
The analysis reveals that the North American banking system is the most influential, causing the largest effects on the other systems. However, feedback effects are  measured with significant causal relations also in the opposite directions. The results are demonstrated to be robust with respect to changes in the method used to compute the transfer entropy, changes in the values of parameters, and with respect to the use of daily or weekly returns in the analysis.

To summarize, the contribution of this paper is three folds.
First, we improve technical aspect of the VAR estimation, allowing for better identification of events concentrated at specific times, which leads to more accurate and insightful interpretation of the results.
Second, we focus on connectedness in banking sector, while previous studies based on the Diebold and Yilmaz methodology analyzed networks of financial institutions. In particular, we analyze the North American, EU and ASEAN banking systems individually and show that, despite the regions' geographical distances, they are affected in various degrees by major financial crisis events originated in dominant regions such as the North American and EU banking systems.
Third, we originally perform a causality analysis on the regional connectedness time series generated through the Diebold and Yilmaz's method.
Our analysis suggests that a regional disaggregated investigation has the advantage of introducing a predictive component to this methodology. 
While the network total connectedness measure identifies increase in regional uncertainty associated with major events that shake the markets, the causality relation between total connectedness in different regions, introduced in this paper, provides a quantitative characterization of the flow of uncertainty form region to region, that could be interpreted as the result of contagion. To the best of our knowledge, this causality analysis is the first of its kind.

As future directions we will compare this approach with with other information theoretic measures with the aim to find a framework that is capable to qualify financial uncertainty and its causal effects at all levels of aggregation, from a local single-variable perspective to the global world-market view.

\section*{Acknowledgements}
T.A. and F.C. acknowledge support of the UK Economic and Social Research Council (ESRC) in funding the Systemic Risk Centre (ES/K002309/1). 

\bibliographystyle{jae}
\renewcommand{\bibname}{References}

\newpage

\appendix

\section{List and summary statistics of banks in the sample}
\label{appendix:list of banks}
\begin{table}[ht]
\caption{List of banks that are headquartered in North America (Canada and the U.S.) and have actively traded between 2005-2015}
\label{table:banks-na}
\begin{center}
\begin{footnotesize}
\begin{tabular}{rlrrr}
\toprule 
& Bank name & Country & Daily mean return (\%) & Daily volatility (\%)\\
\midrule 
1. & Canadian Imperial Bank (CIBC) & CAN & 0.01 & 1.82 \\
2. & Bank of Montreal (BMO) & CAN & 0.01 & 1.69 \\
3. & Royal Bank of Canada (RBC) & CAN & 0.03 & 1.73 \\
4. & Toronto Dominion Bank (TD) & CAN & 0.03 & 1.65 \\
5. & Bank of Nova Scotia (BNS) & CAN & 0.01 & 1.72 \\
6. & Citigroup (CITI) & USA & -0.08 & 3.70 \\
7. & Bank of America Corp (BAC) & USA & -0.04 & 3.51 \\
8. & Wells Fargo \& Co (WFC) & USA & 0.02 & 2.86 \\
9. & JP Morgan Chase \& Co (JPM) & USA & 0.02 & 2.64 \\
10. & US Bancorp (USB) & USA  & 0.01 & 2.32 \\
\bottomrule
\end{tabular}
\end{footnotesize}
\end{center}
\end{table}  

\label{table:list of ASEAN banks}
\begin{table}[H]
\caption[List of actively traded ASEAN banks]{List of banks that are headquartered in Southeast Asia and have actively traded between 2005-2015.} 
\label{table:List of ASEAN banks} 
\begin{center}
\begin{footnotesize}
\begin{tabular}{p{0.5cm} p{3.9cm} p{1cm} R{1.5cm} R{1.4cm} R{1.2cm}}
\toprule
& Bank & Country                     & Market cap (\$ billion) & Average return (\%) & Volatility (\%)\\ \midrule
1    & Bank Rakyat Indonesia       & IDN        & 20.43                    & 0.07            & 2.56 \\
2    & Bank Permata                & IDN        & 0.54                    & 0.02            & 1.93 \\
3    & Bank Danamon                & IDN        & 2.23                     & 0.00            & 2.73 \\
4    & Bank Maybank Indonesia      & IDN        & 0.79                    & 0.00            & 2.67 \\
5    & Bank Cimb Niaga         & IDN        & 1.07                   & 0.02            & 2.51 \\
6    & Panin Bank                  & IDN        & 0.17                   & 0.03            & 2.68 \\
7    & Bank Negara Indonesia       & IDN        & 6.66                 & 0.04            & 2.50 \\
8    & Bank Central Asia           & IDN        & 23.21                 & 0.08            & 2.06 \\
9    & Bank Mandiri                & IDN        & 15.75                 & 0.05            & 2.54 \\
\hline
10   & Public Bank                 & MYS        & 16.15                & 0.04            & 0.90 \\
11   & Malayan Banking             & MYS        & 18.70                 & 0.00            & 1.23 \\
12   & RHB Capital                 & MYS        & 3.73               & 0.03            & 1.58 \\
13   & AMMB Holdings               & MYS        & 3.04                & 0.01            & 1.51 \\
14   & AFFIN Holdings              & MYS        & 0.97                    & 0.01            & 1.65 \\
15   & Alliance Financial Group    & MYS        & 1.15                 & 0.01            & 1.52 \\
16   & BIMB Holdings               & MYS        & 1.35                    & 0.03            & 2.13 \\
17   & CIMB Group Holdings         & MYS        & 7.92                & 0.02            & 1.54 \\
18   & Hong Leong Bank             & MYS        & 6.17                & 0.03            & 1.14 \\
\hline
19   & Philippine National Bank    & PHL        & 1.20                    & 0.03            & 2.39 \\
20   & Bank of Philippine Islands  & PHL        & 6.97                   & 0.03            & 1.79 \\
21   & China Banking Corp          & PHL        & 1.36                    & 0.04            & 1.39 \\
22   & Metropolitan Bank and Trust & PHL        & 4.67          & 0.05            & 2.12 \\
23   & Security Bank Corp          & PHL        & 1.86                   & 0.07            & 1.87 \\
24   & Rizal Commercial Bank Corp  & PHL        & 0.94                    & 0.03            & 2.19 \\
25   & Union Bank                  & PHL        & 1.22                    & 0.05            & 1.77 \\
26   & BDO Unibank                 & PHL        & 7.33                   & 0.05            & 2.04 \\
\hline
27   & United Overseas Bank    & SGP        & 19.62                   & 0.01            & 1.49 \\
28   & DBS Group Holdings      & SGP        & 25.23                   & 0.01            & 1.49 \\
29   & Oversea-Chinese Banking     & SGP        & 22.71                    & 0.02            & 1.33 \\
\hline
30   & Krung Thai Bank             & THA        & 6.79                    & 0.02            & 2.11 \\
31   & Siam Commercial Bank        & THA        & 11.44                    & 0.03            & 2.02 \\
32   & Bangkok Bank                & THA        & 8.04                    & 0.02            & 1.81 \\
33   & Bank of Ayudhya             & THA        &  6.15                   & 0.03            & 2.41 \\
34   & Kasikornbank                & THA        & 10.94                   & 0.04            & 1.97 \\
35   & TMB Bank                    & THA        & 3.12                    & -0.01           & 2.40 \\
36   & Kiatnakin Bank              & THA        & 0.91                    & 0.00            & 1.94 \\
37   & Tisco Financial Group       & THA        & 0.96                    & 0.02            & 2.11 \\
38   & Thanachart Capital          & THA        & 14.3                    & 0.03            & 2.13 \\
39   & CIMB Thai Bank              & THA        &  0.76                  & -0.01           & 2.75 \\ \bottomrule
\end{tabular}
\end{footnotesize}
\end{center}
\end{table}

\begin{table}[]
\begin{center}
\begin{footnotesize}
\caption{List of banks that are headquartered in the EU and have actively traded between 2005-2015 (1).}
\label{table: EU banks 1}
\begin{tabular}{@{}lllll@{}}
\toprule
& Bank & Country                       & Daily return (\%) & Volatility (\%)      \\ \midrule
1    & Oberbank Ag                   & AUT               & 0.02            & 0.38 \\
2    & Erste Group Bk Ag             & AUT               & -0.01           & 2.95 \\
3    & KBC Group Nv                  & BEL               & 0.00            & 3.50 \\
4    & Dexia Sa                      & BEL               & -0.21           & 7.76 \\
5    & Hellenic Bank                 & CYP               & -0.08           & 3.08 \\
6    & Komercni Banka As             & CZE               & 0.01            & 2.10 \\
7    & IKB Deutsche Industriebank    & DEU               & -0.13           & 3.90 \\
8    & Commerzbank                   & DEU               & -0.08           & 3.09 \\
9    & DVB Bank Ag                   & DEU               & 0.03            & 1.38 \\
10   & HSBC Trinkaus \& Burkhardt    & DEU               & 0.00            & 1.73 \\
11   & Comdirect Bank Ag             & DEU               & 0.02            & 1.83 \\
12   & Deutsche Postbank Ag          & DEU               & 0.00            & 2.15 \\
13   & Danske Bank As                & DNK               & 0.01            & 2.11 \\
14   & Jyske Bank                    & DNK               & 0.02            & 1.94 \\
15   & Nordea Invest Fjernosten      & DNK               & 0.01            & 1.43 \\
16   & Sydbank As                    & DNK               & 0.03            & 1.93 \\
17   & Banco Santander Sa            & ESP               & 0.00            & 2.16 \\
18   & BBVA                          & ESP               & -0.01           & 2.12 \\
19   & Banco Popular Espanol         & ESP               & -0.07           & 2.30 \\
20   & Bankinter                     & ESP               & 0.01            & 2.28 \\
21   & Banco De Sabadell Sa          & ESP               & -0.02           & 1.89 \\
22   & BNP Paribas                   & FRA               & 0.00            & 2.56 \\
23   & Natixis                       & FRA               & -0.01           & 3.12 \\
24   & Societe Generale Group        & FRA               & -0.02           & 2.86 \\
25   & Credit Agricole Sa            & FRA               & -0.02           & 2.78 \\
26   & CIC (Credit Industriel Comm)  & FRA               & 0.00            & 1.41 \\
27   & Barclays Plc                  & GBR               & -0.03           & 3.23 \\
28   & HSBC Hldgs Plc                & GBR               & -0.02           & 1.72 \\
29   & Royal Bank of Scotland Group  & GBR               & -0.10           & 3.91 \\
30   & Standard Chartered Plc        & GBR               & 0.00            & 2.44 \\
31   & Lloyds Banking Group Plc      & GBR               & -0.05           & 3.37 \\
32   & Piraeus Bank Sa               & GRC               & -0.22           & 5.04 \\
33   & Attica Bank Sa                & GRC               & -0.23           & 5.88 \\ \bottomrule
\end{tabular}
\end{footnotesize}
\end{center}
\end{table}
\begin{table}[]
\begin{center}
\begin{footnotesize}
\caption{(cont.) List of banks that are headquartered in the EU and have actively traded between 2005-2015 (2).}
\label{table: EU banks 2}
\begin{tabular}{@{}lllll@{}}
\toprule
& Bank & Country                       & Daily return (\%) & Volatility (\%)      \\ \midrule
34   & Eurobank Ergasias Sa          & GRC               & -0.31           & 5.52 \\
35   & National Bank of Greece       & GRC               & -0.20           & 4.81 \\
36   & Alpha Bank Sa                 & GRC               & -0.15           & 4.69 \\
37   & Zagrebacka Banka              & HRV               & 0.00            & 2.58 \\
38   & Privredna Banka Zagreb Dd     & HRV               & 0.01            & 2.37 \\
39   & OTP Bank Plc                  & HUN               & 0.00            & 2.63 \\
40   & Unicredit Spa                 & ITA               & -0.05           & 2.90 \\
41   & Credito Emiliano Spa          & ITA               & 0.00            & 2.26 \\
42   & Intesa Sanpaolo Spa           & ITA               & 0.00            & 2.61 \\
43   & Banca Popolare Di Sondrio     & ITA               & -0.01           & 1.83 \\
44   & Banca Carige Spa Gen \& Imper & ITA               & -0.10           & 2.39 \\
45   & Banco Desio Della Brianza     & ITA               & -0.02           & 1.76 \\
46   & Banco Popolare                & ITA               & -0.06           & 2.86 \\
47   & Banca Popolare Di Milano      & ITA               & -0.03           & 2.78 \\
48   & Banca Monte Dei Paschi Siena  & ITA               & -0.12           & 2.96 \\
49   & Bank of Siauliai Ab           & LTU               & -0.06           & 2.97 \\
50   & ING Groep Nv                  & NLD               & -0.01           & 3.14 \\
51   & Van Lanschot Nv               & NLD               & -0.03           & 1.62 \\
52   & Mbank Sa                      & POL               & 0.05            & 2.34 \\
53   & Bank Handlowy W Warzawie Sa   & POL               & 0.01            & 2.05 \\
54   & ING Bank Slaski Sa            & POL               & 0.04            & 1.90 \\
55   & Bank BPH S.A.                 & POL               & -0.09           & 4.48 \\
56   & Bank Millennium Sa            & POL               & 0.03            & 2.62 \\
57   & Bank Plsk Kasa Opk Grp Pekao  & POL               & 0.00            & 2.26 \\
58   & Bank Zachodni Wbk Sa          & POL               & 0.04            & 2.15 \\
59   & Getin Holding Sa              & POL               & -0.02           & 3.16 \\
60   & Powszechna Kasa Oszczednosci  & POL               & 0.00            & 2.02 \\
61   & Banco BPI Sa                  & PRT               & -0.03           & 2.46 \\
62   & Banco Comercial Portugues Sa  & PRT               & -0.09           & 2.76 \\
63   & Svenska Handelsbanken         & SWE               & 0.02            & 1.86 \\
64   & Skandinaviska Enskilda Bank   & SWE               & 0.01            & 2.55 \\
65   & Nordea Bank Ab                & SWE               & 0.02            & 2.05 \\
66   & Swedbank Ab                   & SWE               & 0.01            & 2.53 \\ \bottomrule
\end{tabular}
\end{footnotesize}
\end{center}
\end{table}
\end{document}